\documentclass[conference]{IEEEtran}
\IEEEoverridecommandlockouts

\usepackage[style=ieee]{biblatex}
\usepackage{tikz}
\usetikzlibrary{quantikz}

\usepackage{amsmath,amssymb,amsfonts}
\usepackage{algorithmic}
\usepackage{graphicx}
\usepackage{textcomp}
\usepackage{xcolor}
\def\BibTeX{{\rm B\kern-.05em{\sc i\kern-.025em b}\kern-.08em
    T\kern-.1667em\lower.7ex\hbox{E}\kern-.125emX}}
\usepackage{subcaption}

\newcommand{\comment}[1]{}

\begin{document}

\title{Comparing Quantum Machine Learning Approaches in Astrophysical Signal Detection
\thanks{This research has been partially supported by the ICSC Italian national research center on High Performance Computing, Big Data, and Quantum Computing.   
It has also been conducted in the context of the CTAO consortium, and 
supported by the National Recovery and Resilience Plan (PNRR), Mission 4, Component 2, Investment 1.4, Call for tender No. 1031 published on 17/06/2022 by the Italian Ministry of University and Research (MUR), funded by the European Union – NextGenerationEU, Project Title ``National Centre for HPC, Big Data and Quantum Computing'' (HPC) – Code National Center CN00000013 – CUP D43C22001240001.
}
}

 \author{\IEEEauthorblockN{Mansur Ziiatdinov}
 \IEEEauthorblockA{\textit{MIFT Department} \\
 \textit{University of Messina}\\
 Messina, Italy \\
 mansur.ziiatdinov@unime.it}
 \and
 \IEEEauthorblockN{Farida Farsian}
 \IEEEauthorblockA{\textit{OACT, INAF} \\
 Catania, Italy\\
 farida.farsian@inaf.it}
 \and
 \IEEEauthorblockN{Francesco Schillir\'o}
 \IEEEauthorblockA{\textit{OACT, INAF} \\
 Catania, Italy \\
francesco.schilliro@inaf.it}
\and
\IEEEauthorblockN{Salvatore Distefano}
\IEEEauthorblockA{\textit{MIFT Department} \\
\textit{University of Messina}\\
Messina, Italy \\
salvatore.distefano@unime.it}
}


\maketitle

\begin{abstract}

Machine Learning (ML) serves as a general-purpose, highly adaptable, and versatile framework for investigating complex systems across domains. 
However, the resulting computational resource demands, in terms of the number of parameters and the volume of data required to train ML models, can be high, often prohibitive. 
This is the case in astrophysics, where multimedia space data streams usually have to be analyzed.
In this context, quantum computing emerges as a compelling and promising alternative, offering the potential to address these challenges in a feasible way.
Specifically, a four-step quantum machine learning (QML) workflow is proposed 
encompassing data encoding, quantum circuit design, model training and  evaluation. 
Then, focusing on the data encoding step, different techniques and models are investigated within a case study centered on the Gamma-Ray Bursts (GRB) signal detection 
in the astrophysics domain. 
The results thus obtained demonstrate the effectiveness of QML in astrophysics,  
highlighting the critical role of data encoding,  which significantly affects the QML model performance.
\end{abstract}


\section{Introduction}
\label{intro}


Machine learning (ML) models have proven to be remarkably suitable for 
problems
where traditional, rule-based programming falls short, and the relationships within the data are complex, non-linear and difficult to define explicitly. 
ML models can indeed investigate complex problems, uncovering hidden patterns and making predictions through data, even
noisy or incomplete ones, filtering out outliers to analyze the underlying trends. 
They are also able to adapt to changing environments, learning and evolving as new data become available, thus successfully adopted 
to, e.g., categorize data, predict values and events, group similar items, recommend preferences-based choices, support decision making, detect anomalies, recognize and tracking objects.

ML has emerged as a transformative tool in astrophysics, enabling the analysis of vast, high-dimensional datasets that traditional statistical methods struggle to handle. 
Modern astronomical surveys generate extensive and complex data streams, necessitating automated classification, anomaly detection, and predictive modeling. 
ML techniques have been widely adopted in tasks such as galaxy classification, transient detection, and cosmological parameter estimation, demonstrating significant improvements in accuracy and efficiency \cite{Fluke:2020ml, Baron:2019ml}. 
While ML has proven effective, astrophysical data present additional challenges beyond sheer volume. 
ML, indeed, well-suits astrophysical research due to the confluence of several factors: i) astrophysical datasets are characterized by their high dimensionality and volume, often exceeding the capacity of traditional statistical methods; ii) astrophysical phenomena are frequently governed by intricate physical processes that generate complex, non-Gaussian statistical distributions;
iii) the inherent noise and incompleteness of astronomical observations, arising from instrumental limitations and observational constraints, require robust data imputation, filtering and denoising techniques; iv) the need for automated classification and pattern recognition in large-scale surveys, such as those conducted by modern telescopes, demands efficient and accurate algorithms; 
v) the ability of ML models to learn and adapt to evolving datasets and to perform predictive modeling is crucial for understanding dynamic astrophysical phenomena. 
However, this complexity often makes classical approaches inefficient or suboptimal, prompting the exploration of alternative computational paradigms.

Quantum Machine Learning (QML) leverages the principles of quantum computing such as superposition, entanglement, and interference, to address computational problems that may be intractable for classical algorithms. 
Unlike classical ML, where scalability and feature representation become limiting factors, QML holds promise for solving problems involving complex patterns and high-dimensional data, dealing with 
ML issues like energy efficiency and large dependence on the training data, but its exploration is still far from the end.
Despite its theoretical potential, QML applications in astrophysics remain largely unexplored, with only a few pioneering studies demonstrating its feasibility. 
For example, quantum-enhanced support vector machines have been utilized for galaxy classification \cite{hassanshahi2023quantum}, quantum neural networks have been applied to classify gravitational waveforms \cite{fan2023data}, and quantum kernel methods have been employed for cosmological data analysis \cite{peters2021machine}. 
However, substantial research is still required to refine QML algorithms and fully exploit quantum computing advantages in astrophysical contexts \cite{carleo2019machine}.

One fundamental aspect of QML is data encoding, which transforms classical data into quantum states for processing. 
Effective encoding is critical for leveraging quantum advantages, with techniques such as basis encoding (binary mapping), amplitude encoding (embedding data in state amplitudes), and angle encoding (using rotation angles). 
The choice of encoding directly impacts model efficiency, expressiveness, and scalability, posing challenges such as hardware limitations and data normalization. 
The successful application of QML in astrophysics largely depends on the choice of data encoding, as astrophysical data consist of large amounts of classical information that must be mapped to quantum states.
Therefore, study of different types of data encoding for astrophysical applications is more than crucial.

To such a purpose, this paper 
investigates QML patterns in astrophysics, characterized into a four-step workflow: data encoding, model design, training, and testing.
Then, it explores in detail such steps adopting a quantum circuit model, proposing techniques and solutions to deal with related issues and challenges.
Focusing on data encoding, it proposes and investigates different techniques in QML models for astrophysics, i.e. amplitude, angle and hybrid techniques such as
the quantum fingerprinting  
and the data reuploading ones. 
After designing the corresponding QML circuits, these techniques are compared by a case study on Gamma-Ray Burst detection.
%

This paper extends a preliminary work  \cite{Farsian:2025viw} towards a generalization of the QML  workflow, models, and techniques that can be adopted in the astrophysics domain.
Specifically, the main contribution of this paper is on: i) defining the QML process and workflow; ii) proposing and review different data encoding techniques; iii) specifying the QML circuit models thus resulting; and iv) comparing the proposed QML models through a real case study on Gamma-Ray Burst detection. 

The remainder of the paper is organized as follows: Section \ref{sec:preliminary-concepts} defines the basic concepts for QML in astrophysics, framed into a 4-step workflow, which steps are discussed in the following sections.
Section \ref{sec:data-encoding} details the data encoding, in fact, while Section \ref{sec:circuit-design} and Section \ref{sec:training-testing} discuss the design, training, and testing of the quantum circuits, respectively.
Section \ref{sec:implementation} reports a case study on Gamma-Ray Burst detection, exploited as a testbed and benchmark to test and compare the proposed techniques and models.
Section \ref{sec:conclusions} wraps the paper up with some final remarks and hints for future work.

\section{Preliminary Concepts}\label{sec:preliminary-concepts}

\subsection{Problem Description}
As outlined above,
QML presents a novel approach to analyzing astrophysical data, particularly in scenarios involving complex patterns and multi-dimensional correlations. To demonstrate the application of QML in astrophysics, this study focuses on Gamma-Ray Bursts (GRB), among the most energetic astrophysical phenomena in the universe.
Originating from distant galaxies emitting intense gamma-ray radiation, GRB are categorized into short-duration GRB (lasting \(<\) 2 seconds), generally associated with neutron star mergers, and long-duration GRB (\(>\) 2 seconds), which are linked to the core collapse of massive stars into black holes \cite{2009A_A...496..585G}. 
These luminous transients serve as crucial probes for fundamental physics and cosmology, contributing to the understanding of general relativity, cosmic reionization, and possible Lorentz invariance violations \cite{Fynbo:2006du, 2009astro2010S.284S}. 
This study focuses on the detection and characterization of long-duration GRB.

Significant advancements in GRB studies have been driven by both space- and ground-based observatories. 
The Fermi Gamma-Ray Space Telescope, with its Large Area Telescope (LAT) and Gamma-Ray Burst Monitor, has played a crucial role in GRB detection and spectral analysis \cite{Thompson:2022ufx, 2020ApJ...893...46V, 2021ApJ...913...60P}. 
The NASA Swift Observatory provides precise localization and multi-wavelength follow-up observations \cite{SwiftScience:2004ykd}, while the AGILE satellite, operated by the Italian Space Agency, contributes to the study of transient Gamma-Ray phenomena, including GRB \cite{AGILE:2008nyq, 2022ApJ...925..152U}. 
Future observatories such as the Cherenkov Telescope Array Observatory (CTAO) are expected to revolutionize GRB research by enabling real-time analysis and improving detection sensitivity \cite{CTAConsortium:2017dvg}.

Given the ever-increasing complexity and volume of GRB datasets, there is a need for novel computational techniques capable of efficiently extracting meaningful patterns. Classical ML approaches have proven valuable in this domain, yet their scalability and performance may be limited when dealing with high-dimensional feature spaces. This is where QML presents a compelling alternative, offering potentially more efficient feature representation, enhanced pattern recognition capabilities, and speedup in certain computational tasks. 
In the next section,  the fundamental principles of QML and  its categorization are outlined, highlighting its potential impact on astrophysical data analysis.

\subsection{Quantum Machine Learning}
QML are categorized according to a well-known taxonomy in 3 main categories, combining the type (classical - C, quantum - Q) of data (D) and algorithms (A) adopted, resulting in 3 classes: CDQA, QDCA and QDQA.
The most common approach is CDQA using both classical and quantum data, with classical data often encoded into quantum states and then processed by QML algorithms. 
QML algorithms are typically hybrid, combining quantum circuits for processing with classical computers for tasks like optimization. 
QML resides where classical data is transformed and processed by hybrid algorithms, yielding classical results. 
The key of CDQA is the interplay between classical and quantum, with data encoding bridging the gap.

Quantum Neural Networks (QNN) are computational models that integrate the principles of quantum computing with the architecture of classical neural networks. 
Unlike classical neural networks that process information using bits, QNN utilize qubits and quantum gates to perform computations. 
This quantum approach allows the creation of neural networks that can potentially handle more complex data patterns and achieve higher computational efficiency.
Quantum counterparts of most common classical NN have been implemented.
Quantum Support Vector Machines (QSVM) leverage the power of quantum kernels to map classical data into high-dimensional Hilbert spaces, where linear separation may be more easily achieved. Quantum kernels, which are computed using quantum circuits, can provide exponential speedups in kernel evaluation compared to classical methods. 
Quantum Convolutional Neural Networks (QCNNs) adapt the convolutional architecture of classical CNNs to the quantum domain. QCNNs utilize quantum circuits to perform convolutional operations, enabling the extraction of features from quantum data. 
Other QNN architectures include quantum autoencoders, quantum generative adversarial networks (QGANs), and hybrid quantum-classical neural networks. Each architecture has its own strengths and is suited for specific applications.

\begin{figure}[ht!]
\vspace{-2cm}
    \centering
    \includegraphics[width=.5\columnwidth]{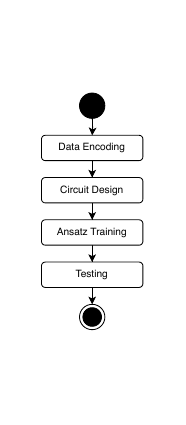}
\vspace{-2cm}
    \caption{VQA PQC Learning Workflow}
    \label{fig:wflow}
\end{figure}

Quantum Neural Network (QNN) solvers, primarily Variational Quantum Algorithms (VQAs), leverage mathematical models
to address machine learning tasks, mainly parameterized quantum circuits (PQC)  and  Quantum Approximate Optimization Algorithms (QAOA)-based adiabatic models are notable examples. 
The circuit model is more flexible, general and widely adopted.
Thus, focusing on the VQA implementation into PQC, the PQC designing and training process can be described by the workflow shown in Fig. \ref{fig:wflow}. 
Assuming that the dataset has been initially processed through a proper feature engineering and optimization  (e.g. by dimensionality reduction) stages performed in the classical domain and resulting into numerical feature encoding (i.e. numbers), it is composed of 4 main steps: Data Encoding, circuit design, ansatz training and testing.
First, classical data are transformed into quantum states through data encoding, a critical step that specifies how information is represented in the quantum system. 
Different techniques for quantum data encoding encompass basis, amplitude, angle, and other high-level encodings, even hybrid ones, affecting the QML model performance and resilience \cite{sharma2024survey,ranga2024quantum,agliardi2025quantum,sierra2023data,bhabhatsatam2023hybrid}. 
Second, the circuit design phase involves selecting an appropriate ansatz, or structure, for the PQC. 
The ansatz choice is crucial, as it determines the circuit expressivity and trainability. 
QML often employs hybrid models where quantum circuits are parameterized and optimized using classical methods requiring parameter initialization \cite{ardeshir2024hybrid} and automation \cite{gomez2022towards}. 
Third, during circuit training, the PQC parameters are optimized using classical optimization algorithms to minimize a cost function quantifying the model performance. 
This is typically done using gradient descent methods tailored for quantum circuits, such as the REsource Frugal Optimizer for QUantum Stochastic gradient descent (Refoqus), which helps in reducing the computational overhead \cite{moussa2023resource}.
The hybrid approach may be effective, e.g. by updating the parameters on classical devices while the quantum circuit processes the data \cite{bhabhatsatam2023hybrid,ardeshir2024hybrid}
This iterative optimization process refines the circuit parameters until a satisfactory level of accuracy is achieved. 
Finally, the trained circuit is tested on testing data to evaluate its generalization performance, ensuring the model effectiveness in real-world applications \cite{sureshquantifying}.
The generalization error is analyzed to ensure the model robustness \cite{caro2022generalization}.

\subsection{ML and QML in Astrophysics}
Classical and quantum machine learning (ML) have been widely adopted in astrophysical domains. 
Classical ML methodologies, encompassing random forests, support vector machines, and neural networks, allow
the detection of extrasolar planets through the analysis of extensive telescopic datasets 
enable the classification of transient celestial objects and quasars, 
aid in signal discrimination from instrumental artifacts 
and predict Solar activities 
\cite{Fluke2020}.

While classical ML effectively recognizes patterns, it may struggle with highly intricate patterns. 
Astrophysics and cosmology have seen emerging applications of QML. 
For instance, quantum-enhanced support vector machines have been successfully applied to galaxy classification, demonstrating the ability to process high-dimensional astronomical datasets \cite{Hassanshahi:2023xgv}. 
QML has also been exploited in radio astronomy, where a quantum neural network was developed for pulsar classification, leveraging quantum-specific encoding methods to achieve comparable accuracy to classical techniques \cite{Kordzanganeh:2021azm}).
The integration of quantum computing with classical ML paradigms potentially yields computational advantages in processing substantial astrophysical datasets 
improving model parallelism and fidelity \cite{Mishra2021, Buffoni2020}. 
Quantum neural networks are shown to be effective in classifying events within high-energy physics, a methodology that is extensible to astrophysical data analysis, achieving comparable or even better results than classical ML with reduced resource demands \cite{Cugini2023}. 
The inherent quantum properties of superposition, interference, and parallelism may enable superior pattern recognition within complex and large astrophysical datasets \cite{Mahim2023, Mishra2021}.
Emergent applications include hybrid models, such as quantum-enhanced kernels in support vector machines, even applied to astrophysical time-series classification \cite{Baker2024}. 
Data augmentation techniques, like conditional generative adversarial networks (cGANs), can generate balanced datasets for ML training, addressing data imbalance in astrophysical observations \cite{Dubenskaya2024}.

Both classical and quantum ML present transformative potential in astrophysics. 
Classical ML is currently prevalent across diverse applications. 
QML is an emerging tool that could offer significant computational advantages and enhanced pattern recognition capabilities. 
The synergistic application of these technologies may advance astrophysical understanding through more efficient and precise data analysis.

To unlock this potential, a systematic and methodological approach, i.e. the one  
proposed in Fig. \ref{fig:wflow},  must be properly implemented and adapted to the astrophysics domain.
For such a purpose, this paper proposes different techniques for data encoding and models for their analysis, assessing them through a real case study in the astrophysics domain.

\section{Data Encoding}\label{sec:data-encoding}
The starting point of the workflow of \figurename \ref{fig:wflow} focuses on data processing, i.e. the quantum encoding.
More specifically, a dataset $\mathbf{X}$ is a set of $n$ $f-$dimensional tuples 
$$x_i=( x_{i,1},...,x_{i,f})\in \mathbf{X} \subseteq \mathbb{R}^{f}$$
where $i=1,..,n$ identifies a single element of the dataset from $n$, that is, the total number of items or the dataset size.
While $x_{i,j} \in \mathbb{R}$ is the $j^{th}$ feature value of the $i^{th}$ data item.
If necessary, an appropriate feature engineering step should also be used to extract categorical features into numerical ones.
As stated above, different quantum encoding techniques can be adopted to obtain a quantum representation, e.g. basis-, amplitude- or angle-based ones, strongly depending on the dataset $\mathbf{X}$ size $n$ and number of features $f$, resulting in a $q$-qubit representation.
In the following, some of the most relevant ones are described.


\subsection{Amplitude Encoding}\label{sec:amplitude-data}

Consider a dataset $\mathbf{X}$,
the amplitude encoding converts the dataset to a state
\[
  \ket{\psi(\mathbf{X})} = \frac{1}{\|\mathbf{X}\|}\sum_{i=1}^{n}\sum_{j=1}^{f} x_{i,j} \ket{i}\ket{j},
\]
where \(\|\mathbf{X}\| = \sqrt{\sum_{i=1}^{n}\sum_{j=1}^{f} x_{i,j}^{2}}\) is an \(\ell_{2}\)-norm of the dataset \(\mathbf{X}\) considered as a vector.

\subsection{Angle Encoding}\label{sec:angle-encoding}

The angle encoding~\cite{schuld2021_mlbook} represents a data item \(x_{i} = (x_{i,1}, \dots, x_{i,f}) \in \mathbb{R}^{f}\) using different qubit rotations, for example, \(R_{Y}\):
\[
  \ket{\psi(x_{i})} = \prod_{j=1}^{f} R_{Y}^{(j)}(x_{i,j})\ket{0^{\otimes f}},
\]
where \(R_{Y}^{(j)}(\theta)\) is a rotation of \(j\)-th qubit about Y-axis by the angle \(\theta\).

\subsection{Data Reuploading}\label{sec:reuploading-data}

The data reuploading encoding uses single-qubit gates to encode classical features and repeats this process interspersing encoding and some computation~\cite{PerezSalinas2020}. 
The motivation behind the data reuploading method is to circumvent the No-Cloning theorem, and to allow the quantum neural network take the same input several times. 
Usually, one layer of data reuploading uses single-qubit gates to encode the features and then entangles all qubits, so, in some sense, it is a generalization of the angle encoding since it applies this technique in different basis. 
Moreover, this method, in principle, could be used as a universal classifier \cite{PerezSalinas2020}.

Unfortunately, it is hard to write the closed-form analytical expression for a data reuploading encoded state. 
Schuld et al.~\cite{Schuld_2021a} show that it can be written as a partial Fourier series
\[
  \ket{\psi(x_{i})} = \sum_{\omega \in \Omega} c_{\omega} e^{i\omega \cdot x_{i}} \ket{0},
\]
where \(\omega \cdot x\) is the inner product, the frequency spectrum \(\Omega \subset \mathbb{R}^{f}\) depends only on the eigenvalues of the data encoding gates, and the  \(c_{\omega}\) coefficients are determined by the design of the circuit.

It is easier to describe data reuploading encoding in terms of transformations that construct the state \(\psi(x_i)\). 
Consider a data item \(x_i = (x_{i,1}, \dots, x_{i,f})\) and choose a number of qubits \(q\) and a number of layers \(l\) such that \(f = lq\).
The state is then obtained by using the following transformation:
\begin{equation}\label{eq:data-reuploading-formula}
\begin{aligned}
  \ket{\psi(x_{i})}
  &= \prod_{k=l}^{1} U_{l-k+1}(x_{i,f-kq+1}, x_{i,f-kq+2} \dots, x_{i,f-kq+q}) \ket{0} \\
  &= U_{l}(x_{i,f-q+1}, \dots, x_{i,f}) U_{l-1}(x_{i,f-2q+1}, \dots, x_{i,f-q}) \times\\
  &\times \dots \times U_{1}(x_{i,1}, \dots, x_{i,q}) \ket{0},
\end{aligned}
\end{equation}
where \(U_{k}(z_{0}, \dots, z_{q-1})\) describes the \(k\)-th layer:
\[
  U_{k}(z_{0}, \dots, z_{q-1}) = \Big( \otimes_{j=1}^{q} R_{Y}^{(j)}(z_{j}) \Big) V_{k},
\]
and \(R_{Y}^{j}(z)\) is a rotation about Y-axis by the angle \(z\) and \(V_{k}\) is a unitary transformation that entangles all qubits (it may vary for different layers).

\subsection{Quantum Fingerprinting  Encoding}\label{sec:qfp-data}
The quantum fingerprinting is a technique that compresses a long classical input into a much shorter quantum register while preserving the essential properties of the input, allowing different inputs to be distinguished. 
It was introduced by Buhrman et al.~\cite{Buhrman2001} to solve the EQUALITY problem in the simultaneous message passing model (i.e.,\ Alice and Bob have their inps ts \(x\) and \(y\) correspondingly, and simultaneously send one message to the referee who computes the output). 
It was then developed and applied to different problems, such as quantum cryptography 
\cite{Ablayev2016a}, quantum string algorithms~\cite{Ablayev_2024}, etc.
Two types of data encoding are considered based on the quantum fingerprinting technique~\cite{Ablayev2016a}.  
Starting from \(\theta = (\theta_{0}, \dots, \theta_{f-1}) \in (0;2\pi)^{f}\) coefficients,  quantum fingerprinting encoding for a data item \(x_i = (x_{i,1}, \dots, x_{i,f}) \in \mathbb{R}^{f}\) can be defined as
\begin{equation}\label{eq:trainable-kernel}
  \ket{\psi_{\theta}(x_i)} = \frac{1}{\sqrt{f}} \sum_{j=1}^{f} \ket{j} \Big( \cos(\theta_{j}x_{i,j})\ket{0} + \sin(\theta_{j}x_{i,j})\ket{1} \Big).
\end{equation}
In other words, different subspaces (indexed by \(j\)) are adopted to encode features \(x_{i,j}\) of the data item \(x_i\) using different rotations \(\theta_{j}\) for each feature. 
This puts the quantum fingerprinting encoding in between the amplitude and angle encodings.
Let refer to the encoding \eqref{eq:trainable-kernel} as FP encoding if some optimization technique is used to search for the optimal coefficient values \(\theta\).
As an alternative,
suggested by the quantum fingerprinting theory, these values can be randomly assigned. 
In this case, the resulting encoding, identified as the FP-Random encoding, can be expressed by:
\begin{equation}\label{eq:simple-kernel}
  \ket{\psi(x_i)} = \frac{1}{\sqrt{f}} \sum_{j=1}^{f} \ket{j} \Big( \cos(\theta_{j}x_{i,j})\ket{0} + \sin(\theta_{j}x_{i,j})\ket{1} \Big),
\end{equation}
where the coefficients \(\theta \in (0;2\pi)^{f}\) are randomly initialized.

\subsection{Comparison}

\begin{table}[htbp]
\caption{Data encoding comparison.}
\begin{center}
\begin{tabular}{|l|c|c|p{3cm}|}
\hline
  \textbf{Encoding}
  &\textbf{Pars}
  &\textbf{\(q\)}
  &\textbf{Description}\\
\hline
  Amplitude & $n,f$ & \(\lceil \log (nf) \rceil\) & Encodes the whole dataset \\
  \hline
  Angle & $f$ & \(f\) & Feature-based encoding\\
\hline
  Data reuploading & $f,l\) & \(\lceil f / l \rceil\) & Mixes amplitude and angle encodings into \(l\) layers \\
\hline  
  Fingerprinting & $f$ & \(\lceil \log f \rceil + 1\) & Mixes amplitude and angle encodings\\
\hline
\end{tabular}
\label{tab:encoding-comparison}
\end{center}
\end{table}
The encoding techniques above discussed are summarized and compared in \tablename~\ref{tab:encoding-comparison} in terms of the number of qubits $q$ based on relevant parameters, i.e. the dataset size\(n\), the number of features \(f\), and the number of levels $l$ (for data reuploading). 
This comparison highlights that the best encoding techniques, in terms of $q$, are the hybrid ones, i.e. data reuploading and fingerprinting.
However, for a deeper comparison in the QML realm is required to investigate the QML performance such as training time, accuracy and similar statistics as done in Section \ref{sec:implementation}.


\section{Circuit Design}\label{sec:circuit-design}

\begin{figure}[htbp]
  \centering
  \begin{subfigure}{0.5\textwidth}
    \centering
    \begin{quantikz}
\lstick{$|0\rangle$} & \gate[3]{S(x_i)} & \gate[3]{S^\dagger(x_j)} & \meter{} \\
\dots                &                &                       & \qw\dots \\
\lstick{$|0\rangle$} &                &                       & \meter{}
    \end{quantikz}
    \caption{Quantum kernel \(k(x_{i},x_{j})\). The gate \(S(x_{i})\) encodes the data item \(x_{i}\).}
    \label{fig:high-level-kernel}
  \end{subfigure}
  \begin{subfigure}{0.5\textwidth}
    \centering
    \begin{quantikz}
\lstick{$|0\rangle$} & \gate[3]{S(x_i)} & \gate[3]{W(\theta)} & \meter{} \\
\dots                &                &                     & \qw\dots \\
\lstick{$|0\rangle$} &                &                     & \meter{}
    \end{quantikz}
    \caption{Quantum neural network. The gate \(S(x_{i})\) encodes the input data \(x_{i}\) and the gate \(W(\theta)\) is the ansatz.}
    \label{fig:high-level-qnn}
  \end{subfigure}
  \caption{High-level structure of QML circuits}
  \label{fig:high-level-circuits}
\end{figure}
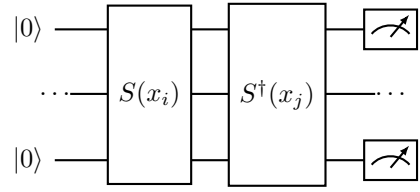
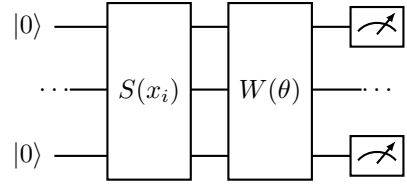

The next step in the workflow of \figurename \ref{fig:wflow} is to design the PQC including quantum kernels and quantum neural networks. 
Their high-level structure is depicted in Fig.~\ref{fig:high-level-circuits}. 
The quantum kernel depicted in \figurename~\ref{fig:high-level-kernel} first uses a data encoding gate \(S(x_{i})\) to prepare the state \(\ket{\psi(x_{i})} = S(x_{i})\ket{0^{\otimes q}}\) and then uses the inverse gate \(S^{\dagger}(x_{j})\) to compute the kernel \(\kappa(x_{i},x_{j}) = \braket{\psi(x_{j})}{\psi(x_{i})}\).
The quantum neural network in \figurename~\ref{fig:high-level-qnn} encodes the data \(x_{i}\) in a quantum state and searches for the optimal parameters \(\theta\) of the ansatz \(W(\theta)\).
The role of the ansatz is to reduce the search space from the set of all unitary transformations of \(q\) qubits to the set of unitary transformations parameterized by \(\theta\), which has a simpler description.


The following subsections focus on the data encoding circuits (also called feature maps in the QNN context) and the ansatz used in the following case study.

\subsection{Data Encoding Circuit}

\subsubsection{Data Reuploading}\label{sec:reuploading-circuit}

\begin{figure}[htbp]
  \centering
  \begin{subfigure}{0.5\textwidth}
    \centering
    \begin{quantikz}
\qw & \gate{R_Y(z_0)}    & \gate[5]{V_{k}}& \qw \\
\qw & \gate{R_Y(z_1)}    &                & \qw \\
\qw & \gate{R_Y(z_2)}    &                & \qw \\
\wave&                   &                &     \\
\qw & \gate{R_Y(z_{q-1})}&                & \qw
    \end{quantikz}
    \caption{The structure of one layer \(U_{k}(z_{0}, \dots, z_{q-1})\) of the Data Reuploading circuit.}
    \label{fig:data-reuploading-circuit-main}
  \end{subfigure}
  \begin{subfigure}{0.5\textwidth}
    \centering
    \begin{quantikz}
\qw & \ctrl{1}& \qw     & \qw\dots\qw & \qw     & \ctrl{5}& \qw \\
\qw & \targ{} & \ctrl{1}& \qw\dots\qw & \qw     & \qw     & \qw \\
\qw & \qw     & \targ{} & \qw\dots\qw & \qw     & \qw     & \qw \\
\wave&        &         &             &         &         &     \\
\qw & \qw     & \qw     & \qw\dots\qw & \ctrl{1}& \qw     & \qw \\
\qw & \qw     & \qw     & \qw\dots\qw & \targ{} & \targ{} & \qw
    \end{quantikz}
    \caption{The qubit entanglement transformation \(V_{\mathrm{odd}}\) for odd layers.}
    \label{fig:entanglement-odd}
  \end{subfigure}
  \begin{subfigure}{0.5\textwidth}
    \centering
    \begin{quantikz}
\qw & \targ{}  & \qw      & \qw\dots\qw & \qw      & \targ{}  & \qw \\
\qw & \ctrl{-1}& \targ{}  & \qw\dots\qw & \qw      & \qw      & \qw \\
\qw & \qw      & \ctrl{-1}& \qw\dots\qw & \qw      & \qw      & \qw \\
\wave&         &          &             &          &          &     \\
\qw & \qw      & \qw      & \qw\dots\qw & \targ{}  & \qw      & \qw \\
\qw & \qw      & \qw      & \qw\dots\qw & \ctrl{-1}& \ctrl{-5}& \qw
    \end{quantikz}
    \caption{The qubit entanglement transformation \(V_{\mathrm{even}}\) for even layers.}
    \label{fig:entanglement-even}
  \end{subfigure}
  \caption{Data Reuploading Circuit}
  \label{fig:data-reuploading-circuit}
\end{figure}
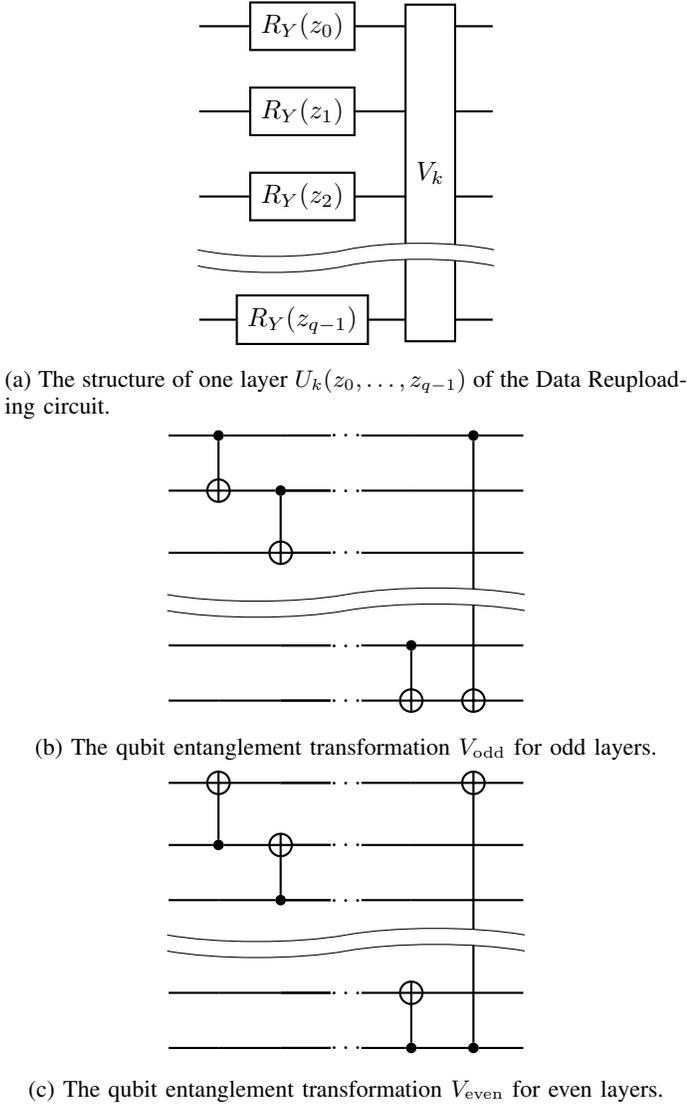
The circuit implementation of the data reuploading method is 
based on the Equation~\eqref{eq:data-reuploading-formula} describing the encoding process, where only the qubit entanglement transformation \(V_{k}\) has to be specified. 
The circuit for encoding a data item \(x_{i} = (x_{i,1}, \dots, x_{i,f}) \in \mathbb{R}^{f}\), where \(f = lq\), is depicted in \figurename~\ref{fig:data-reuploading-circuit-main}.

Different qubit entanglement procedures \(V_{\mathrm{odd}}\) and \(V_{\mathrm{even}}\) for odd and even layers of the encoding, respectively, are exploited: \(V_{1} = V_{\mathrm{odd}}\), \(V_{2} = V_{\mathrm{even}}\), \(V_{3} = V_{\mathrm{odd}}\), \dots, \(V_{l} = V_{\mathrm{odd/even}}\), as depicted
in \figurename s~\ref{fig:entanglement-odd} and~\ref{fig:entanglement-even}.

\subsubsection{Quantum Fingerprinting  Encoding}\label{sec:qfp-circuit}

\begin{figure}[htbp]
  \centering
  \begin{quantikz}
\lstick{$\ket{0}$} & \gate{H} & \ctrl{4}   & \ctrl{4}   & \ctrl{4}   & \qw\ldots & \octrl{4}  & \qw \\
\wave&&&&&&&& \\
\lstick{$\ket{0}$} & \gate{H} & \ctrl{2}   & \ctrl{2}   & \octrl{2}  & \qw\ldots & \octrl{2}  & \qw \\
\lstick{$\ket{0}$} & \gate{H} & \ctrl{1}   & \octrl{1}  & \ctrl{1}   & \qw\ldots & \octrl{1}  & \qw \\
\lstick{$\ket{0}$} & \qw      & \gate{R_1} & \gate{R_2} & \gate{R_3} & \qw\ldots & \gate{R_d} & \qw
  \end{quantikz}
  \caption{Fingerprinting circuit with parameters \(\theta\) for encoding the input \(x_{i}\). Gate \(R_{k} = R_{X}(\theta_{k}x_{i,k})\) rotate the last qubit around X-axis by the angle \(\theta_{k}x_{i,k}\).}
  \label{fig:fingerprinting-circuit}
\end{figure}
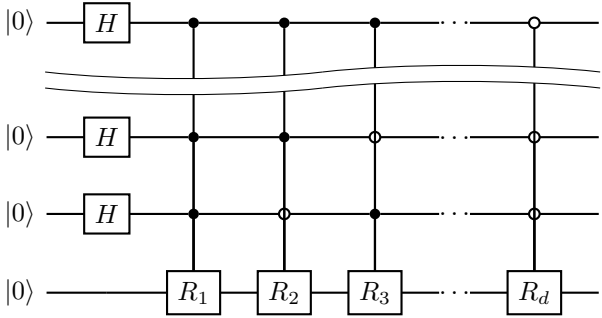

There are different approaches to implement quantum fingerprinting techniques by, e.g., 
optimizing the model to different hardware architectures~\cite{Zinnatullin2023,khadiev2025}, or adopting mathematical methods~\cite{Ziiatdinov_2023} to decrease the depth of the quantum fingerprinting circuit. 
Since such experiments focus on 
data encoding and the noiseless state vector simulation is used, these advanced approaches are not required and the quantum fingerprinting data encoding is implemented in a naïve way, by performing controlled rotations in different subspaces as shown in \figurename~\ref{fig:fingerprinting-circuit}.

\subsection{Variational Quantum Circuit-Ansatz}\label{sec:ansatz}

Different ansatz implementations are available in literature, 
customized for the hardware 
or the internal structure of the task. 
In the astrophysics context, the ansatz of~\cite{Sukin_2019} is selected for its suitability to astrophysical problems, as discussed
in 
~\cite{Farsian:2025viw}.
The ansatz \(W(\theta)\) is defined by the same components of data reuploading: \(W(\theta) = U_{2}(\theta_{q}, \dots, \theta_{2q-1}) U_{1}(\theta_{0}, \dots, \theta_{q-1})\), where the layers \(U_{k}(z_{0}, \dots, z_{q-1})\) have been defined above (see Equation~\eqref{eq:data-reuploading-formula} and \figurename~\ref{fig:data-reuploading-circuit}).
Therefore, the ansatz has \(2q\) parameters, where \(q\) is the number of qubits.

\section{Training and Testing}\label{sec:training-testing}

QML model training aims at finding the optimal parameter values to minimize the loss function: after this, the model can represent, with a given accuracy, the data points from the same distribution of the training data.
Each QML model specifies its own parameters and loss function, while
the testing phase is the same for the QNN and quantum kernel methods.

The whole data set \(\mathbf{X}\) of \(|\mathbf{X}| = n\) items is divided into two non-intersecting subsets, the training data set \(\mathbf{X}_{tr}\) and the test data set \(\mathbf{X}_{test}\) with \(|\mathbf{X}| = n_{tr}\) and \(|\mathbf{X}_{test}| = n_{test}\) data items, respectively. The training data set is used to fit the model parameters, and the test data set is used only for assessment.

For the classification problem, each data item \(x_{i} \in \mathbf{X}\) is supplemented with a label \(y_{i} \in \{1,\dots,c\}\), where \(c\) is the number of classes. The goal of a classification model is to correctly predict the label of previously unseen data items.

\subsection{Quantum neural networks}\label{sec:qnn}

For quantum neural networks the loss function is defined as the cross-entropy between the target distribution \(\{y_{i}\}\) and the measured results \(\{\hat{y}_{i}\}\):
\[
  \ell = -\sum_{i=1}^{n} y_{i} \log \hat{y}_{i}.
\]
To compute gradients of the loss function the parameter shift rule~\cite{Wierichs2022} can be exploited, minimizing it by the classical gradient descent methods, such as the Constrained Optimization By Linear Approximation (COBYLA) from the scipy package~\cite{2020SciPy-NMeth}.

\subsection{Quantum kernel methods}\label{sec:q-kernel}

The quantum kernel method implements the same kernel trick as the classical kernel method: it may be easier to distinguish data items if mapped into a space of higher dimension (in the case of the quantum kernel method, this space is a Hilbert space).
Thus, for a data set \(\{(x_i, y_i)\}\), the quantum kernel method optimizes the weights \(w_i\) of data items and uses the kernel \(k(x_i, x')\) to predict the result for a new data item \(x'\), where \(x' \in \mathbb{R}^{f}\).
For example, for binary classification, the quantum kernel method defines the following estimate:
\begin{equation}\label{eq:kernel-estimate}
\hat{y} = \mathrm{sgn} \sum_{i=1}^n w_iy_ik(x_i, x'),
\end{equation}
where weights are usually trained using support vector machines (SVM).

The essence of the quantum kernel method is to compute the kernel function on a quantum computer, possibly leveraging the available computational advantage. 
To such a purpose, a sample \(x_i\) is encoded in a state \(\ket{\psi(x_i)}\), and the kernel is computed as the inner product of two states: \(k(x_{i},x_{j}) = \langle \psi(x_{j}) | \psi(x_{i}) \rangle\).

In real experiments, the QSVC class from the qiskit-machine-learning library\footnote{Qiskit Machine Learning: \url{https://qiskit-community.github.io/qiskit-machine-learning/}} can be used.
Inferring the model by Equation \eqref{eq:kernel-estimate} implies evaluating the quantum kernel \(n\) times, and the loss function requires assessing the kernel \(k(x_{i}, x_{j})\) for all pairs of data points (i.e. \(n(n-1)/2\) times).
The PegasosQSVC class is based on the Pegasos (Primal Estimated sub-GrAdient SOlver) algorithm~\cite{Shalev-Shwartz2011} to estimate the loss function by random sampling.
If the kernel \(k_{\theta}(x,y)\) has customizable parameters \(\theta\), these are optimized before training the SVM itself by, for example, using the SPSA (Simultaneous Perturbation Stochastic Approximation)~\cite{spall1998-spsa} optimizer.

\subsection{Assessment}\label{sec:assessment}

Standard metrics like accuracy and \(F_{1}\) score (see, e.g.,~\cite{muller2016}) are exploited for the assessment of classification tasks. 
For each model, the number \(TP, TN, FP, FN\) of true positive, true negative, false positive, and false negative predictions, respectively, are collected. 
The \textit{accuracy} $a$ is defined as
\[
  a = \frac{TP + TN}{TP + TN + FP + FN}.
\]
The \(F_{1}\) \textit{score} is defined as the harmonic mean of \textit{precision} \(\mathrm{prec}\) and \textit{recall} \(\mathrm{recall}\):
\[
  F_{1} = 2 \cdot \frac{\mathrm{prec} \cdot \mathrm{recall}}{\mathrm{prec} + \mathrm{recall}}
\]
where \(\mathrm{prec} = TP / (TP + FP)\) and \(\mathrm{recall} = TP / (TP + FN)\).

\section{Case Study}\label{sec:implementation}

\subsection{Dataset and Simulation Framework}\label{sec:dataset}

As introduced in Section \ref{sec:preliminary-concepts}, this case study focuses on the detection of GRB signals. To achieve this, a simulated dataset that accurately represents both GRB events and background noise is required. 
This dataset serves as the foundation for evaluating the performance of different QML models.  
To generate synthetic GRB datasets, \textit{Gammapy}, a widely used Python package for high-level gamma-ray data analysis \cite{gammapy:2023}, has been adopted. 
Gammapy supports simulations based on the Gamma-Ray Astronomy Data Format (GADF) and incorporates instrument response functions (IRFs) essential for accurate astrophysical modeling \cite{universe7100374, gammapy:2024}.  

The dataset comprises 600 simulated light curves, split into 480 training samples and 120 test samples in the standard case. The dataset is balanced, with half containing GRB signals and the other half representing background noise. The simulation is configured as follows:  

\textbf{Source Model}: The GRB is modeled as a transient point-like gamma-ray source positioned 0.4° off-axis, a typical observational configuration for very-high-energy studies. The spectral model follows a power-law distribution with an index of 2.25, resembling a Crab-like source. 
Given the ongoing uncertainty regarding prompt TeV gamma-ray emissions from GRBs \cite{2022icrc.confE.998P}, a generic power-law spectrum is adopted rather than attempting a detailed spectral reconstruction. 
The source flux normalization is set to \(3 \times 10^{-10} \mathrm{cm^{-2} s^{-1} TeV^{-1}}\) and scaled by a uniformly sampled factor between 0.1 and 3.0 to introduce variability.  

\textbf{Temporal Model}: The light curves are shaped using a Gaussian pulse model, representing a first-order approximation of observed GRB prompt emissions at MeV energies. While real GRB light curves often consist of multiple overlapping pulses, a simplified single-pulse approach is adopted to facilitate training. Pulse durations and peak times are randomized, ensuring that the peak remains within the central 800 s of the 1200 s time window to capture both the rise and decay phases.  

\textbf{Simulated Data Products}: The simulations generate event lists recording individual photon detections over time, which are subsequently binned into count maps. The total duration of each event list is 1200 s, with adjustable binning strategies for analysis.  

\textbf{Instrument Response Functions (IRF)}: The Prod5 IRF for the CTAO Large-Sized Telescope (LST) array are exploited, assuming a 20° zenith angle. 
The LST, with their wide field of view and low-energy sensitivity, are well-suited for transient gamma-ray observations \cite{cherenkov_telescope_array_observatory_2021_5499840}. 
\begin{figure}[ht]
    \centering
    \includegraphics[width=1\linewidth]{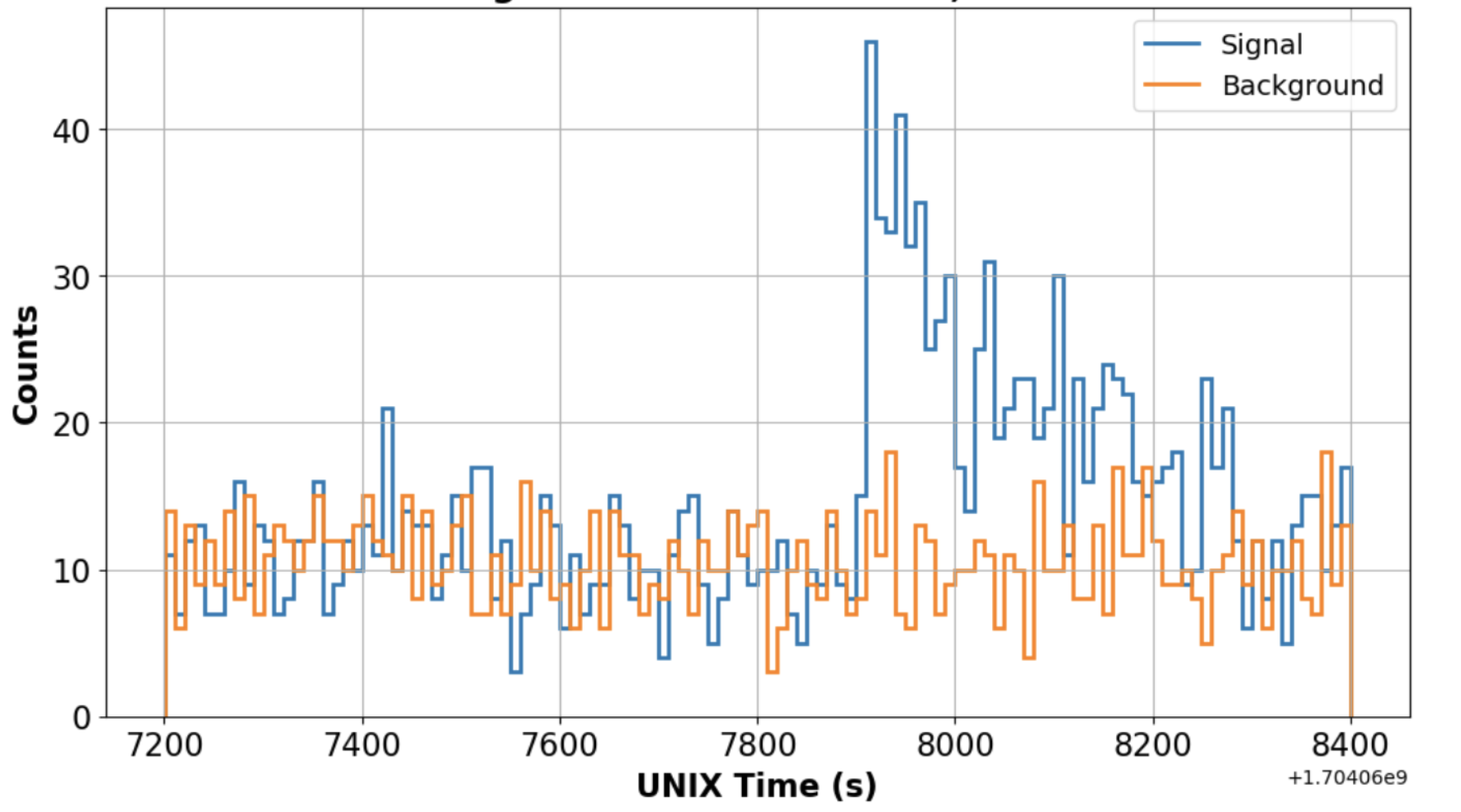}
    \caption{An example of simulated light curves used in training and test set. Blue curve shows the GRB signal with underlying background while orange shows only the background noise. This curve was sampled with bin=10s.}
    \label{fig:LC}
\end{figure}

Following the simulation, the light curves are binned into time intervals of 5, 10, 50, and 100 seconds, corresponding to feature dimensions of 240, 120, 24, and 12, respectively. 
These different binning strategies allow for a comparative analysis of model performance across varying temporal resolutions.
\figurename \ref{fig:LC} presents an example of a simulated GRB light curve. The x-axis represents time in 10s bins, while the y-axis denotes photon counts per bin. 
The blue curve illustrates the GRB signal overlaid with background noise, whereas the orange curve represents background-only events.
For more detailed information about the dataset, available on request, see \cite{Farsian:2025viw}.

\subsection{Experiments, Results and Comparison}\label{sec:comparison}

The experiments are designed to provide insights into how different data encoding techniques affect the training and inference performance of quantum machine learning in the detection of GRB signals. 
This way, different encoding techniques have been investigated 
to shed light on the trade-offs between accuracy, efficiency, and resource requirements. 

The following sections detail the testbed for the experiments, the hyperparameter setup and, 
once identified an optimal QML model configuration, a comparison of classical and quantum ML through the GRB dataset, further investigating on fingerprinting (FP) and data reuploading (DR) encoding techniques.

\subsubsection{Test environment}\label{sec:testbed}

All tests were performed on a machine with the following characteristics:
\begin{itemize}
  \item CPU:\@ AMD Ryzen 9 5950X
  \item RAM:\@ 64 GiB
  \item OS:\@ Linux, kernel: 6.6.74-gentoo
  \item Python: 3.12.8
  \item jupyter-core: 5.7.2
  \item numpy: 2.2.2
  \item qiskit: 1.3.2
  \item qiskit-machine-learning: 0.8.2
  \item scikit-learn: 1.6.1
\end{itemize}
The source code is available on request.

\subsubsection{Hyperparameter Setup}
\label{sec:pegasos-c}

\begin{table}[htbp]
\caption{FP data encoding parameter setup experiments} 
\begin{center}
\begin{tabular}{|c|c|c|c|c|c|c|c|c|}
\hline
  {Bin}
  &{Enc.}
  &{Opt.}
  &$f$ 
  &$q$ 
  &$n_{tr}$ 
  &{Alg.}
  &{$a$} 
  &{Time}\\ 
\hline
  100 & FP & yes & 12 & 5 & 480 & QSVC & 97.5\% & \(\approx 10\) h\\
  50 & FP & yes & 24 & 6 & 480 & QSVC & 96.7\% & \(\approx 20\) h\\
  100 & FP-R & no & 12 & 5 & 480 & QSVC & 97.5\% & \(\approx 2.5\) h\\
  100 & FP-R & no & 12 & 5 & 480 & Pegasos & 97.5\% & \(\approx 2.5\) m\\
\hline
\end{tabular}
\label{tab:preliminary-experiments}
\end{center}
\end{table}

Since the quantum fingerprinting (FP) method provides a degree of freedom in the selection of the training process hyperparameters, while the data reuploading (DR) method is well-known, the first initial experiments focused on identifying the 
quantum fingerprinting data encoding parameters.
The results thus obtained (see \tablename~\ref{tab:preliminary-experiments}) show that the FP encoding parameter optimization  process is slow (simulation wall time between 2.5-20 h)
and does not lead to significant improvement in the accuracy against random parameter settings FP-R, where R stands for random. 
The Qiskit SDK provides an alternative classifier algorithm, the Pegasos Quantum Support Vector Classifier (Pegasos)~\cite{Shalev-Shwartz2011}, which  provides better performance in experiments (2.5 m).
Thereby, in the rest of the experiments only FP-R and data reuploading (DR) encodings are considered, and thus
FP  will refer to the Pegasos random fingerprinting encoding. 

Furthermore, the Quantum Support Vector Classifiers (QSVC)  kernel training  requires \(\approx n_{tr}^2\) kernel computations (where \(n_{tr}\) is the training dataset size). 
The Pegasos algorithm chooses random subsets to estimate the kernel loss function, and therefore its running time does not depend on the size of the training dataset \(n_{tr}\).
The regularization hyperparameter \(C\) allows to tune this process. 
Larger values of \(C\) may lead to overfitting, but make some steps of the algorithm trivial and thus faster.
The $C$ setup experiments are conducted on the $f=12$ feature GRB dataset, split into 480 samples for training and 120 for testing. 
To mitigate the noise effects of the FP
kernel random initialization, 10 runs of the algorithm with the same train/test split and different random parameters \(\theta\) are performed. 
Each simulation run consists of 500 iterations of the Pegasos QSVC algorithm; at the end of the run the accuracy $a$, $F_{1}$ score, and the training wall time are evaluated.

\begin{table}[htbp]
  \caption{Pegasos QSVC $C$ experiment results}
\begin{center}
\begin{tabular}{|r|r|r|r|}
\hline
\(C\) & $a^{\mathrm{a}}$ (\%) & $F_1^{\mathrm{a}}$ (\%) & Time $^{\mathrm{a}}$ (s)\\
\hline
1 & 82.0 & 78.0 & 435.97\\
10 & 92.6 & 92.0 & 387.01\\
100 & 96.7 & 96.6 & 150.24\\
1000 & 97.5 & 97.5 & 61.44\\
\hline
\multicolumn{4}{l}{$^{\mathrm{a}}$Median value for 10 runs}
\end{tabular}
\end{center}
\label{tab:pegasos-c-results}
\end{table}

The experiment results are summarized in \tablename~\ref{tab:pegasos-c-results}, confirming that larger values of the regularization hyperparameter \(C\)  make Pegasos faster, also improving 
the accuracy $a$ and the $F_{1}$ score.
Since the results are influenced by outliers, the median value of 10 runs is reported.

\subsubsection{Classical vs Quantum ML}
\label{sec:running-time}

One of the possible advantages of quantum machine learning may be faster computation time. 
Since the models are only simulated and not executed on the real quantum hardware, it is not correct to directly compare the running time of quantum algorithms with their classical counterparts.
To overcome this issue and ensure a fair comparison, an estimate of the time it would take to execute the algorithm on a quantum computer is provided. 
The quantum fingerprinting circuit execution is simulated 100 times and the simulation wall time \(T_{\mathrm{sim}}\) is measured. 
Since the state vector simulation is used, the time 
$$T_{\mathrm{sim}} = N_{\mathrm{reps}}\cdot d \cdot t_{\mathrm{sim,1}}$$ 
is a product of the repetition number \(N_{\mathrm{reps}}\) (due to taking multiple shots for measurement and repeating the execution several times), the depth of the circuit \(d\), and the time for simulating a single gate \(t_{\mathrm{sim,1}}\). 
To estimate the time to run the same circuit on a real quantum hardware, the single gate simulation time is replaced with the time for executing the gate on a quantum computer \(t_{\mathrm{real,1}} = 50\mathrm{ ns}\) (IQM~\cite{abdurakhimov2024_iqm} reports 20--40 ns), thus 

$$T_{\mathrm{real}} = N_{\mathrm{reps}}\cdot d \cdot t_{\mathrm{real,1}} = T_{\mathrm{sim}} \cdot t_{\mathrm{real,1}} / t_{\mathrm{sim,1}}.$$

Thereby, the experimental setup for comparing the running time of quantum and classical machine learning method is the following. 
The GRB dataset with 12 features and different number of training samples for different experiments is used to find the dependence on the number of samples. 
For the same training data, 10 random initializations of FP-Random kernel are fitted with Pegasos algorithm (with \(C = 1000\) and 100 iterations) and the results are compared with the classical RBF kernel. 
The results of the experiments are summarized in \tablename~\ref{tab:running-time}.

\begin{table}[htbp]
  \caption{Classical (RBF) and quantum (FP, DR) ML experiments results with $f=12$}
\begin{center}
\begin{tabular}{|l|r|r|r|r|r|}
  \hline
Model$^{\mathrm{a}}$ & $n_{tr}$ & $a^{\mathrm{b}}$ & $F_1^{\mathrm{b}}$ & $T_{\mathrm{sim}}^{\mathrm{b}}$ (ms)& 
$T_{\mathrm{real}}$ (ms)\\
\hline

RBF &  6 & 97.5 & 97.4 &           & 0.451\\
FP &   6 & 91.7 & 90.9 &  3482.751 & 0.322\\
DR &   6 & 50.0 & 66.4 &  1886.430 & 0.174\\
  \hline
RBF & 12 & 96.8 & 96.6 &           & 0.457\\
FP &  12 & 97.3 & 97.2 & 11475.969 & 1.061\\
DR &  12 & 50.0 & 66.5 &  4370.821 & 0.404\\
  \hline
RBF & 30 & 97.2 & 97.1 &           & 0.472\\
FP &  30 & 96.8 & 96.7 & 28850.884 & 2.667\\
DR &  30 & 50.0 & 66.4 & 10688.307 & 0.988\\
  \hline
RBF & 60 & 97.2 & 97.1 &           & 0.494\\
FP &  60 & 96.8 & 96.7 & 29863.474 & 2.760\\
DR &  60 & 51.8 & 67.2 & 17006.314 & 1.572\\
  \hline
\multicolumn{6}{l}{$^{\mathrm{a}}$RBF classical, FP  fingerprinting, DR data reuploading}\\
\multicolumn{6}{l}{$^{\mathrm{b}}$Median value for 10 runs}\\
\end{tabular}
\end{center}
  \label{tab:running-time}
\end{table}

While the simulation time $T_{sim}$ for quantum methods is much higher than the running time $T_{real}$ for classical one, the comparison among $T_{real}$  assuming quantum methods run on quantum computer, shows that these times are, in fact, similar, and further development of the quantum hardware may even lead to the advantage of the quantum machine learning. 
Specifically, the quantum method $T_{real}$ value is smaller than the classical one up to \(n_{tr}=12\). 
The accuracy $a$ and \(F_{1}\) score of classical RBF and quantum FP kernels are almost equal starting from \(n_{tr}=12\), while quantum DR encoding does not learn if the training set size is lower than \(n_{tr}\leq 60\).

\subsubsection{Data Reuploading vs Fingerprinting}
\label{sec:dr-fp-comp}

Finally, data reuploading and fingerprinting encoding methods are compared on the GRB dataset
considering 12, 24, and 120 features for GRB detection through QNN.
To such a purpose, the GRB dataset has been split into a training dataset with $n_{tr}=480$ items and a testing one with $n_{test}=120$ items.
The number of features $f$ determines the number of qubits $q$ for the quantum fingerprinting method (i.e. 5 qubits for 12 features, 6 --- for 24 features, and 8 --- for 120 features).
For a fair comparison, the number of qubits $q$  of data reuploading experiments is set as close as possible to the quantum fingerprinting one (i.e. $q=6$ for 12 and 24 features, $q=6$ and $q=10$ for 120 features). 
The data reuploading number of layers \(l= f/(2q)\) is thus obtained by \(q\) and the number of features \(f\).
The ansatz has the same structure (described in Section~\ref{sec:ansatz}) for both methods.
The COBYLA optimizer is used for 150 iterations and the results are averaged over 10 random initializations of the FP encoding.

\begin{table}[htbp]
  \caption{Data reuploading and fingerprinting  experiment results}
\begin{center}
  \begin{tabular}{|r|r|l|r|l|l|l|l|}
    \hline
Enc.$^{\mathrm{a}}$ & $f$ & $q$ &  \(l\) & $a^{\mathrm{b}}$ (\%) & $F_1^{\mathrm{b}}$ (\%) & $T_{\mathrm{sim}}^{\mathrm{b}}$ (s) & $T_{\mathrm{real}}$ (ms)\\
\hline
FP & 12 & \textbf{5} & - & \textbf{97.5} & \textbf{97.5} & 223.0 & 20.6\\
DR &12 & 6 &  1 & 66.1 & 75.3 & \textbf{161.4} & \textbf{14.9}\\
\hline
 FP & 24 & 6 &- & 95.0 & 95.1 & 433.5 & 40.1\\
DR &24 & 6 &  2 & \textbf{97.5} & \textbf{97.5} & \textbf{214.7} & \textbf{19.8}\\
\hline
FP &120 & 8 &  - & \textbf{97.5} & \textbf{97.5} & 5365.4 & 495.9\\
DR &120 & \textbf{6} &  10 & 87.5 & 88.0 & \textbf{662.0} & \textbf{61.2}\\
DR & 120 & 10 &  6 & 69.5 & 76.9 & 1110.4 & 102.6 \\
    \hline
\multicolumn{6}{l}{$^{\mathrm{a}}$FP  fingerprinting, DR data reuploading}\\
\multicolumn{6}{l}{$^{\mathrm{b}}$Median value of 10 runs}
\end{tabular}
\end{center}
  \label{tab:dr-fp-conparison}
\end{table}

The results of the experiments are reported in  \tablename~\ref{tab:dr-fp-conparison}. 
They show that the fingerprinting encoding FP can achieve better accuracy $a$ but it is slower than DR. 
As expected, the data reuploading method has some flexibility in the number of qubits, while fingerprinting requires \(\lceil \log_2 n_\mathrm{features} \rceil + 1\) qubits as discussed above.

\section{Conclusions}\label{sec:conclusions}
In conclusion, this research has demonstrated the potential of Quantum Machine Learning (QML) to address the complex challenges inherent in astrophysical data analysis. 
By establishing a comprehensive four-step workflow encompassing data encoding, model design, training, and testing, different QML techniques have been explored and compared, particularly focusing on data encoding techniques such as data reuploading and quantum fingerprinting. 
The comparative study on Gamma-Ray Burst detection effectively showcased the practical application, suitability and effectiveness of these techniques in astrophysics, extending a preliminary work towards a more generalized QML framework. 
It implements a structured approach to QML, including different data encoding methods and resulting QML circuit models,  validated through a real-world case study on Gamma-Ray Burst detection. 
Experiments and results confirm that, while classical machine learning has undeniably revolutionized astrophysical data processing, QML holds the promise of further advancements by leveraging quantum computing capabilities. 
This work highlights the critical role of data encoding in bridging the gap between classical astrophysical data and quantum processing. 
Future research should focus on refining QML algorithms, exploring novel encoding techniques, and addressing the scalability challenges to fully harness the potential of quantum computing application in astrophysics. 

\printbibliography

\end{document}